\documentclass{elsart}
\usepackage{graphicx}
\usepackage{amssymb}

\begin{document}

\begin{frontmatter}

\title{Calculation of Correlation Functions in Terms of Fluctuation 
Relation}

\author{Takaaki Monnai, Shuichi Tasaki}
\ead{monnai@suou.waseda.jp}

\address{Department of Applied Physics ,Waseda University,3-4-1 Okubo,
Shinjuku-ku,Tokyo 169-8555,Japan }

\begin{abstract}
We derive a relation similar to the fluctuation theorem for work 
done on a system obeying Langevin dynamics with thermal and colored 
noises. Then, we propose a method of calculating the correlation 
function 
of the colored noise by using this fluctuation relation.

\end{abstract}

\end{frontmatter}

\section{Introduction}
Fluctuation theorems (FT)~\cite{EvansCohen} have been studied as one of 
exact relations valid in far
 from equilibrium regime.
They state the symmetry of fluctuation of 
entropy production $\Delta S$:
\begin{equation}
\left(\lim_{t\rightarrow \infty}\right)\frac{\pi_t(\Delta S =A)}
{\pi_t(\Delta S = -
A)}= A \ ,
\end{equation}
where $\pi_t(\Delta S=A)$ denotes the probability density that a 
system's entropy production during
time interval $t$ equals to $A$, and 
the limit of $t\to +\infty$ 
is taken 
or not depending on  systems.
Although interpretations and systems differ, this expression 
universally holds~\cite{EvansSearles0,Evans,KurchanCl,Spohn1999,Gallavotti1,Jarzynski}.

Jarzynski equality~\cite{JarzynskiEq}
was shown to be derived from FT~\cite{12} and was applied, e.g. to 
evaluate
the free energy of biological systems~\cite{Liphardt}.
It is even more interesting to investigate direct applications of FT.
In this letter, we derive a FT like relation (FR) for work done on a 
system obeying Langevin dynamics.
In addition to the thermal noise, a colored noise is assumed to be 
applied.
Then, this FR is shown to
be used to evaluate a correlation function 
of the colored noise 
from the 
fluctuations of the work done.
Such correlation functions would provide information 
about driving force in nonequilibrium regime
and we
expect that the FR-aided analysis is useful to investigate 
the behaviors of single molecular motors~\cite{Nishiyama} or micro/nano 
machines. 

\section{Fluctuation Relation and Correlation Function}

We consider a system described by a Langevin equation:
\begin{eqnarray}
\int_{-\infty}^t \Gamma(t-t') v(t') dt'=&&f+\xi(t)+\Xi(t)
\label{Langevin} \\ 
\langle\xi(t)\xi(t')\rangle =&&\frac{1}{\beta}\Gamma(t-t')\label{FDT} 
\\ 
\langle\Xi(t)\Xi(t')\rangle =&&M(t-t') \\
\langle\xi(t)\rangle =&&\langle \Xi(t)\rangle =0 \ , \label{mean}
\end{eqnarray}
where $\xi(t)$ and $\Xi(t)$ are the thermal and colored noises, 
respectively,
$f$ stands for a constant external force, and $\beta$ is the inverse 
temperature. The noises $\xi(t)$ and $\Xi(t)$ are assumed to be Gaussian
and independent from each other. 
Eq.(\ref{FDT}) is nothing but the fluctuation dissipation relation.
We assume that the memory kernel $\Gamma(t)$ exponentially decays as 
$\Gamma(t)=\Gamma_0 e^{-\gamma \mid t\mid}$.
Such a Langevin equation would provide a phenomenological description 
of the 
motion of molecular motors as pointed 
out by Harada~\cite{Harada}, where the energy balance in the molecular 
motor 
was investigated.

Now, we introduce the work $\Sigma_t$ done on the system by the 
external force $f$ 
during the time interval $t$:
\begin{equation}
\Sigma_t \equiv \beta \int_0^t ds f v(s)
\end{equation} 
We denote the probability density of $\Sigma_t$ being $A$ as 
$\pi_t(\Sigma_t=A)$.
As we shall show later, one then obtains a fluctuation relation (FR):
\begin{eqnarray}
F(t)\equiv &&t \left(\left\{\frac{1}{A}\log \frac{\pi_t(\Sigma_t=A)}
{\pi_t(\Sigma_t=-A)}\right\}^{-1}-1\right) \nonumber \\
=&&\frac{\beta}{2 \gamma \Gamma_0 }\int_0^t ds \int_0^t ds'\frac{1}
{2\pi}\int_{-\infty}^{\infty}d\omega (\gamma^2+\Gamma_0^2){\hat M}
(\omega)
e^{i \omega (s-s')}  \ , \label{FT}
\end{eqnarray} 
where ${\hat M}(\omega)$ is the Fourier transformation of the 
correlation 
function $M(t)$ of the colored noise.
Note that this FR is different from usual FT's, because it says nothing 
about entropy production, but only refers to the work done by an 
external force.

The FR (\ref{FT}) leads to a differential equation of $M(t)$:
\begin{equation}
 -\frac{d^2}{dt^2}M(t)+\gamma^2 M(t)=\frac{1}{2\pi}\int_{-\infty}
^{\infty}d\omega(\gamma^2+\Gamma_0^2){\hat M}(\omega)e^{i \omega t}
=\frac{\gamma \Gamma_0}{\beta}\frac{d^2}{dt^2}F(t) \ . \label{Cal}
\end{equation}
The quantity $F(t)$ defined in (\ref{FT}) can be measured 
experimentally by 
observing individual trajectories and, thus, one can determine $M(t)$
provided two parameters $\gamma$ and $\Gamma_0$ are known.
We note that $F(t)$ is independent of $A$.

So far we have assumed $\langle \Xi (t) \rangle =0$ for simplicity. 
However, 
the case of $\langle \Xi (t)\rangle\equiv \Xi_0 \not=0$ can be treated 
in the same way and one can show that the function 
$$
\tilde{F}(t)\equiv t 
\left(\left\{\frac{1}{A}\log \frac{\pi_t(\Sigma_t=A)}{\pi_t(\Sigma_t=-
A)}\right\}^{-1}
-\frac{f}{f+\langle \Xi_0 \rangle} \right)
$$
instead of $F(t)$ satisfies 
\begin{equation}
 -\frac{d^2}{dt^2}M(t)+\gamma^2 M(t)=\frac{\gamma 
(f+\Xi_0 )\Gamma_0}{\beta f}\frac{d^2}{dt^2}\tilde{F}(t) \ .
\end{equation}
In this case again, the correlation function $M(t)$ of the colored 
noise 
can be evaluated experimentally.

\section{Derivation of Fluctuation Relation} 

In this section, we shall derive (\ref{FT}).
Because $\Sigma_t$ is a Gaussian process, in order to determine 
$\pi_t(\Sigma_t=A)$, it is sufficient to
calculate the mean value $\langle \Sigma_t \rangle$, and the variance 
$\langle\Sigma_t^2\rangle-\langle
\Sigma_t\rangle^2$.

With the aid of the Fourier transformation, one obtains following 
expression :
\begin{equation}
v(t)=\frac{1}{2\pi}\int_{-\infty}^{\infty} d\omega\frac{{\hat f}(\omega)
+{\hat \xi}(\omega)+{\hat \Xi}(\omega)}
{{\hat \Gamma}(\omega)}e^{i\omega t},
\end{equation}
where ${\hat \Gamma}(\omega)\equiv \frac{\Gamma_0}{\gamma + i \omega}$ 
is the 
Fourier-Laplace transformation of $\Gamma(t)$, and ${\hat f}(\omega)
\equiv 2
\pi f \delta (\omega)$, ${\hat \xi}(\omega)$ and ${\hat \Xi}(\omega)$ 
are the Fourier 
transformations of $f$, $\xi(t)$, and $\Xi(t)$, respectively.
Therefore, (\ref{mean}) leads to
\begin{equation}
\langle\Sigma_t\rangle=\beta f^2 \frac{\gamma}{\Gamma_0}t
\end{equation}
From $\langle {\hat \Xi}(\omega){\hat \Xi}(\omega')\rangle = 2\pi 
{\hat M}(\omega) \delta (\omega +\omega')$ and 
the fluctuation dissipation theorem: $\langle {\hat \xi}(\omega)
{\hat \xi}(\omega')\rangle=\frac{4 \pi 
\Re{\hat \Gamma}(\omega)\delta(\omega+\omega')}{\beta}$, 
we obtain the variance of $\Sigma_t$:
\begin{eqnarray}
&&\langle\Sigma_t^2\rangle-\langle
\Sigma_t\rangle^2=\beta\frac{2 \gamma f^2}{\Gamma_0} t \nonumber \\
+&&\beta^2 f^2\int_0^t ds \int_0^t ds'\frac{1}{(2\pi)^2}\int_{-\infty}
^{\infty}d\omega\int_{-\infty}^
{\infty}d\omega\int_{-\infty}^{\infty}d\omega'\frac{\langle{\hat \Xi}
(\omega)
{\hat \Xi}(\omega')\rangle}{{\hat \Gamma}(
\omega){\hat \Gamma}(\omega')}e^{i\omega s}e^{i\omega' s'} \\
=&& \beta \frac{2\gamma f^2}{\Gamma_0} t + \beta^2 f^2 \int_0^t ds 
\int_0^t ds' \frac{1}{2\pi}\int
 d\omega \frac{\gamma^2+\omega^2}{\Gamma_0} {\hat M}(\omega) e^{i\omega 
(s-s')} \ .
\end{eqnarray}
Substituting the mean value $\langle\Sigma_t\rangle$ and the variance 
$\langle\Sigma_t^2\rangle-\langle\Sigma_t\rangle^2$ into 
\begin{equation}
\log \frac{\pi_t(\Sigma_t=A)}{\pi_t(\Sigma_t=-A)}
=\frac{2\langle\Sigma_t\rangle A}{\langle\Sigma_t^2\rangle-
\langle\Sigma_t\rangle^2} \ ,
\end{equation} 
one obtains the fluctuation relation for work done by an external 
force: (\ref{FT}).

To summarize, we derived a fluctuation relation for the work done
on the system: eqs.(\ref{FT}) and (\ref{Cal}). 
Eq.(\ref{FT}) has a form of the fluctuation theorem 
with modification due to 
$\langle \Xi(t)\Xi(t')\rangle$.
An important point is that the right hand side of (\ref{Cal}) is 
measurable experimentally, and thus, we 
can evaluate the correlation function $\langle\Xi(t)\Xi(t')\rangle$ of 
the nonthermal colored noise.

\section{Acknowledgement}
The authors are grateful to Professor P.Gaspard, Professor T.Yanagida, 
and Dr. M. Nishiyama for fruitful discussions.
This work is supported by a Grant-in-Aid for
Scientific Research (C) from JSPS, by a Grant-in-Aid for Scientific 
Research of Priority Areas ``Control of Molecules in Intense Laser 
Fields'' 
and the 21st Century COE Program (Holistic research and Education 
Center for
Physics of Self-Organization Systems)
both from the Ministry of Education, Culture, Sports, Science and 
Technology of Japan.

\end{document}